\documentclass[11pt]{article}
\usepackage{amsmath}
\usepackage{graphicx}

\begin{document}
\title{
%Conformal field theory in
%Tomonaga-Luttinger liquid with
%$1/r^\beta$ type long-range interactions
%Application of 
Conformal field theory in
Tomonaga-Luttinger model with
$1/r^\beta$ long-range interaction
}
%\begin{center}
\author{Hitoshi Inoue$^{1,2}$ 
%\footnote{Present address: Department of 
%Theoretical Studies, Institute for Molecular Science,
%Okazaki 444-8585, Japan 
%
%E-mail: hinoue@ims.ac.jp
%Telephone number: 0564-\\
%Fax number:  
%}  
%$^{*}$ Department of Theoretical Studies, Institute for Molecular Science,
%Okazaki 444-8585, Japan
\\
$^1$ Department of Physics, Kyushu University 33, Hakozaki,
Higashi-ku, \\ 
Fukuoka 812-8581, Japan
\\
$^2$
West Komoda 3-14-36, Iizuka 820-0017, Fukuoka, Japan
\footnote{corresponding
phone num.:011-81-948-22-0711}}
\date{\today}
\maketitle
\begin{abstract}
%\abstract
I attempt to construct $U(1)$ conformal field theory(CFT) in the Tomonaga-Luttinger(TL) liquid
with $1/r^\beta$ long-range interaction (LRI).
Treating the long-range forward scattering as a perturbation and applying CFT 
to it,
I derive the finite size scalings which 
depend on the power of the LRI. The obtained finite size
scalings give the nontrivial behaviours when $\beta$ is odd and is close to 2.
%And from finite size corrections in the ground state, 
%I find the effective central charge deviates from $c=1$. 
I find the consistency between the analytical arguments and numerical results in
the finite size scaling of energy.
%Moreover I find that the necessary condition 
%for the CFfT holds in the $1/r^\beta$ type long-range interactions.
\end{abstract}
%\pacs{75.10.-b,75.10.-w,75.10.jm}
\section{Introduction}
Electron systems have attracted our much attention 
in the low energy physics.
As the dimension of the electron systems decrease,
the charge screening effects become less important. 
In spite of these facts, 
models with short-range interaction 
have been adopted in many researches 
of one dimensional electron systems.  
The recent advanced technology makes it possible 
to fabricate quasi-one-dimension systems. Actually 
in low temperature
the effect of Coulomb force has been observed  
in $GaAs$ quantum wires \cite{GaAs},
quasi-one-dimensional conductors \cite{1Dcond1,1Dcond2,1Dcond3}
and 1D Carbon nanotubes \cite{ctube1,ctube2,ctube3}. 

The systems with $1/r$ Coulomb repulsive forward scattering
was investigated on the long distance properties 
by bosonization techniques
\cite{Schulz3}. The charge correlation function
decays with the distance as $\exp (- {\rm const. }(\ln x)^{1/2})$ more slowly 
than any power law. 
The momentum distribution function and the density of state does not show
the simple 
power law singular behaviour.
The logarithmic behaviours appear in the power \cite{Wang}. These mean that the 
system is driven to the Wigner crystal which is quite different from the ordinary
TL liquid.
The investigation for the interaction $1/r^{1-\epsilon}$
through the path integral approach \cite{Iucci} reconfirms the slower decaying of the single 
particle Green function 
for $\epsilon=0$ and leads the faster decay 
for $0 \leq \epsilon (\ll 1)$ than any power type.     

The numerical calculation 
in the electron system with the Coulomb interaction shows that 
the larger range of the interaction causes the insulator (charge density wave) to metal 
(metallic Wigner crystal) transition \cite{Maekawa}. In the spinless fermion system,
the convergence of the Luttinger parameters exhibits the quasi-metallic 
behaviour different from the simple TL one \cite{Capponi}.  

As I will discuss below, the forward scattering is irrelevant for $\beta>1$.  
As an instance of 
the effect of the long-range Umklapp scattering,
it was reported that the $1/r^2$ interaction makes the system 
gapless to gapful through the generalized Kosterlitz-Thouless 
transition \cite{Hatsugai}. 

In this paper I discuss CFT in the 
system with LRI.
The basic assumptions of CFT are symmetries of
translation, rotation, scale and special conformal 
transformation. Besides them I assume the 
short-range interaction in the CFT. 
Hence it is a subtle problem whether the CFT can 
describe the system with LRI. 

Of LRIs, up to now,
the solvable models with $1/r^2$ interaction were discussed 
\cite{Haldane2,Sutherland,Sutherland2,Kawakami2}.
With the Bethe ansatz,
the conformal anomaly and the conformal dimensions were calculated and  
the system proved to be described
by $c=1$ CFT. 
%The excitation energies of the $1/r^2$ systems show the agreement with
%the $c=1$ CFT. 
In fact the central charge from 
the specific heat agrees with $c=1$. 
On the other hand, the ground state energy 
is affected by 
the LRI and the periodic nature. The effective central 
charge
% from the finite scaling
%of the ground state energy 
\footnote{I define the ``effective central charge'' by 
$c^{'}=\frac{6 b}{\pi v}$ in the finite size scaling of the 
ground state energy $E_{g}=a L-\frac{b}{L}$. I use the 
word ``effective central charge'' in this sense.}
deviates from $c=1$. 

In general, the CFT for
LRI which breaks the locality, has been left 
as unsettled problem.
%So I discuss the uncertain $1/r^{\beta}$ type interactions.
It is significant to clarify the validity of the CFT to the 
systems with LRI. 
I investigate the tight-binding model with 
$1/r^{\beta}$ interaction as one of such problems.
The low energy effective model consists of 
TL liquid, the long-range
forward scattering and the long-range spatially oscillating Umklapp scattering.
Extending arguments appearing in Ref. \cite{Ludwig} to the TL liquid with the
long-range forward scattering, I derive the finite size scalings.
%From the obtained energy
%scalings dependent on the poIr $\beta$, I find the
%deviation of the effective central charge from $c=1
In the tight-binding model with 
$1/r^{\beta}$ interaction, 
I calculate numerically the size dependences of energy 
and the coefficients of $1/L^y$. 
And I see numerically the relations 
between the velocity, susceptibility
and Drude weight, which CFT requires. 
%Finally I find the consistency in our arguments. 
%After checking them, I discuss the CFT to the TL liquid with 
%the $1/r^\beta (\beta > 1)$ type
%long-range interactions. 
%
%The strategy is as follows. 
%In section* ${\rm I\!I}$ A and B
%I discuss the long wavelength 
%behaviors of the system when I consider the long-range
%forward scatterings by making use of bosonization techniques and the
%renormalization group (RG) method. In the
%section* ${\rm I\!I}$ C        
%I find the corrections caused by such the $\beta >1$ long-range forwards scatterings
%in the finite size 
%scaling of energies.
%In the section* ${\rm I\!I\!I}$, 
%I analyze a tight-binding model with the $\beta >1$ type 
%long-range interactions numerically 
%by considering these corrections 
%in the finite size scaling. 
%I show numerical data of the drude Iight and 
%compressibility and determine the range of TLL in the interactions
%strength space.
%In section* ${\rm I\!V}$ I argue the case $\beta =1$.
%\section{Field theoretical approaches}
\section{Field theoretical approach}
%\subsection*{Theories}
I consider
the following tight-binding Hamiltonian of the interacting 
spinless Fermions:
\begin{eqnarray}
 H &=& -\sum^{L}_{j} (c^{\dagger}_{j} c_{j+1} + {\rm h.c} )+
\frac{g}{2} \sum^{L}_{i\neq j} (\rho_{i}-1/2)
V(i-j) (\rho_{j}-1/2) \label{Hamillng},
\end{eqnarray}
where the operator $c_{j}$ ($c_{j}^{\dagger}$) 
annihilates (creates) the spinless Fermion in the site $j$ and
$\rho_{j}=c^{\dagger}_{j}c_{j}$
is the density operator. 
In order to treat this model under the periodic boundary condition, 
I define the chord distance between 
the sites $i$ and $j$: 
$r_{i,j}=(\frac{L}{\pi} \sin \frac{\pi(i-j)}{L})$ where 
$L$ is the site number. 
%This chord distance approaches to the 
Using this, I express the LRI as 
$V(i-j)= \frac{1}{(\frac{L}{\pi} \sin \frac{\pi(|i-j|)}{L})^\beta}$. 

By the bosonization technique, I obtain
the effective action of the Hamiltonian
(\ref{Hamillng}) for the arbitrary filling:
\begin{eqnarray}
 S &=& \int d\tau dx \frac{1}{2 \pi K} 
(\nabla \phi)^2 + g\int d\tau dx dx^{'}
\partial_{x} \phi(x,\tau)V(x-x^{'})\partial_{x^{'}} \phi(x^{'},\tau)
\nonumber \\
   &+&g^{'} \int d\tau dx dx^{'}
\cos(2k_{F}x+{\sqrt 2}\phi(x,\tau))V(x-x^{'})
\cos(2k_{F}x^{'}+{\sqrt 2}\phi(x^{'},\tau)), 
\label{actotal}
\end{eqnarray}
where $V(x)=\frac{1}{|x|^\beta}$,
$K$ is the TL parameter and $k_{F}$ is the 
Fermi wave number. And $g^{'}$ is the couping constant proportional
to $g$. 
The first term of (\ref{actotal}) 
is the TL liquid and
the second term is the long-range forward scattering.
The last term is the spatially oscillating Umklapp process which includes
$\cos {2 \sqrt 2} \phi$ which comes from the interaction between the neighbour sites.

Schulz analyzed the effects of
the Coulomb forward scattering by the bosonization 
technique in the electron system\cite{Schulz3}. He discussed the
quasi-Wigner crystal of electrons due to the Coulomb forward scattering.   
Here I focus on the effects of
%what
%happens in considering 
the $1/r^{\beta}$ forward scattering
in the spinless Fermions system.
I treat the action
%\footnote{This action is known for Kibble's model \cite{Kibble}.
%In three dimension, the interactions
%induce the mass gap when $\beta =1$. 
%The situations in one dimension 
%may be different from that in three dimension.}
:
\begin{eqnarray}
 S &=& \int d\tau dx \frac{1}{2 \pi K} (\nabla \phi)^2 
+ g\int d\tau dx dx^{'}
\partial_{x} \phi(x,\tau)V(|x-x^{'}|)
\partial_{x^{'}} \phi(x^{'},\tau) \label{action} 
\end{eqnarray}
for any filling $k_{F}$.
%Precisely speaking, 
To investigate in the Fourier space, 
I choose the form 
$V(x)=\frac{1}{(x^2+\alpha^2)^{\beta /2}}$,
where $\alpha$ is the ultra-violet cut-off.
In the wave number space, the action (\ref{action})
is expressed as
\begin{eqnarray}
S &=& \int dq dw \{ \frac{2 \pi}{K}(q^2+w^2)+ g q^2 V(q) \}|\phi(q,w)|^2
,\label{action2}
\end{eqnarray}
where $V(q)$ is the Fourier transformation of $V(x)$:
\begin{eqnarray}
V(q) &=& \frac{2\sqrt{\pi}}{\Gamma(\beta/2)
2^{\beta/2-1/2}}
(\alpha q)^{\beta/2-1/2} K_{\beta/2-1/2}(\alpha q).
\end{eqnarray}
Here $K_{\nu}(x)$ is the modified Bessel function of $\nu$th order
and $\Gamma(x)$ is the gamma function. 
From this, the dispersion relation is 
\begin{eqnarray}
w^2 &=& q^2\{ 1+\frac{g K }{2 \pi}V(q)\}. 
\label{wgive}
\end{eqnarray}
The long wavelength behaviors of 
$V(q) $ are given by
\begin{equation}
 V(q) \sim
\begin{cases}
%A+B q^{\beta-1} +\cdots  & \text{$ 0< \beta < 1$} \\
%A+B \ln q+\cdots         & \text{$ \beta=1  $} \\
%A+B q^{\beta-1}+\cdots   & \text{$ \beta > 1$} \\
%A+ Bq^{2} \ln q +Cq^{2}+\cdots 
%                         & \text{$ \beta = 3$} \\
%A+Bq^{2}+C q^{\beta-1}\cdots
%                         & \text{$ 5 > \beta > 3$} \\
%A+ B q^{2}+ Cq^{4}\ln q+ Dq^{4}+\cdots 
%                         & \text{$ \beta =5 $} \\
%A+Bq^{2}+C q^{\beta-1} +\cdots
%                         & \text{$ \beta > 5$}, \label{longwave}
A+B(q \alpha ) ^{2}+C (q \alpha )^{\beta-1} +\cdots 
& \text{$ \beta > 0$ and $\beta \neq $ odd} \\
A+B \ln q \alpha +\cdots         & \text{$ \beta=1  $} \\
A+ B(q \alpha)^{2} \ln q \alpha +C(q \alpha)^{2}+\cdots
                         & \text{$ \beta = 3$} \\
A+ B (q \alpha)^{2}+ C(q \alpha)^{4}\ln (q \alpha)
+ D(q \alpha)^{4}+\cdots
                         & \text{$ \beta =5 $} \\
\cdots , 
\label{longwave}
\end{cases}       
\end{equation}
where $A,B,C$ and $D$ are the functions of $\beta$.
For the case where $ \beta > 0$ and $\neq $ odd, 
the coefficient $B=B(\beta),C=C(\beta)$ is given by
\begin{eqnarray}
B(\beta) &=& \frac{\pi^{3/2}}{4}
\frac{1}{2^{\beta/2-1/2} \Gamma(\frac{5-\beta}{2}) \Gamma(\beta/2) 
\sin\frac{(\beta-1)\pi}{2}} \nonumber \\
C(\beta) &=& -\pi^{3/2}
\frac{1}{2^{\beta-1} \Gamma(\frac{\beta+1}{2}) \Gamma(\beta/2) 
\sin \frac{(\beta-1)\pi}{2}}.
\end{eqnarray}
%which take different values for
%respective ranges of $\beta$. 
% ²¼¤Îµ­½Ò¤ÏÀµ¤·¤¤¤È»×¤¦¡ª
%From these behaviors I see that the system is 
%gapless when $\beta > 0$ and it is expected to be the 
%TL liquid when $\beta > 1$, that is, $w \sim q$.
%¤ä¤à¤ò¤¨¤º¼è¤EE¯¤³¤È¤Ë¡£

From the eqs.(\ref{wgive}) and (\ref{longwave}), I see that 
$(q \alpha)^{\beta-1}$ and $\ln q \alpha$ terms
for $0< \beta \leq 1 $ affect the linear dispersion essentially. 
Especially for $\beta =1$, there is the analysis by 
Schulz, where the charge density correlation function is calculated \cite{Schulz3}.
According to it, in the present spinless case, the LRI drives  
the ground state from the TL liquid to the quasi-Wigner crystal
as $\beta \rightarrow 1+$. The slowest decaying part of the density 
correlation function is given by 
\begin{eqnarray}
 <\rho(x) \rho(0) > &\sim& \cos (2k_{F} x) {\exp (-{\rm c} 
{\sqrt {\log x}})},
\end{eqnarray}
where $c$ is a function of K, which exhibits slower spatial decay 
than the power decay of TL liquid.
%\begin{eqnarray}
% <\rho(x) \rho(0) > &\sim& -K \frac{1}{x^2}+ 
%{\rm const.} \frac{\cos (2k_{F} x)}{|x|^K}. 
%\end{eqnarray}

Then I see the effects of the long-range forward 
scattering in the standpoints of 
the renormalization of $g$.
The renormalization group eqs. of $g$, $v$ and $K$ are simply derived
for the long wave-length (see Appendixes.). From the renormalization eqs.,
%\begin{eqnarray}
% \frac{dg}{dl}&=&0, \nonumber \\
% \frac{d}{dl}(\frac{v}{K})&=&\frac{g A}{2\pi}, \qquad 
% \frac{d}{dl}(\frac{1}{vK})=0,
%\end{eqnarray}
%for $\beta =1$, where $A$ is the constant appearing in (\ref{longwave})
%and
%\begin{eqnarray}
%\frac{dg}{dl} &=& (1-\beta) g, \nonumber \\
%\frac{dK}{dl} &=& 0, \qquad \frac{dv}{dl} = 0, \label{RG}
%\end{eqnarray}
%for $0< \beta < 1$ and $1< \beta < 3$. For $\beta=3$ case, 
%I should distinguish the two modes $q^{2}$ and $q^{2} \ln q$
%in the behaviours (\ref{longwave}). For the two modes, I
%need the two couplings $g_{1}$ and $g_{2}$ (the details are given in 
%Appendixes.). 
%The obtained RG eqs. are
%\begin{eqnarray}
%\frac{dg}{dl} &=& -2 g \\
%\frac{d g_{1}(l)}{d l} &=& -2 g_{1}(l), \nonumber \\
%\frac{d g_{2}(l)}{d l} &=& -2 g_{2}(l)-g_{1}(l), \\
%\frac{dK}{dl} &=& 0, \qquad \frac{dv}{dl} = 0. \nonumber
%\end{eqnarray}
%and $g(L)=g(0)(\rm const. \frac{1}{L^{2}} + \rm const.\frac{1}{L^{2}}
%\ln \frac{1}{L})$, where I dare to 
%write the explicit form 
%about $g(L)$ to see easily.
%Moreover I derive 
%\begin{eqnarray}
%\frac{dg}{dl} &=& -2 g, \nonumber \\ 
%\frac{dK}{dl} &=& 0, \qquad \frac{dv}{dl} = 0, \label{RGtype2}
%\end{eqnarray} 
%for $\beta > 3$.
the g terms are relevant for $\beta < 1$, marginal for
$\beta= 1$ and irrelevant for $\beta > 1$. Thus it is expected that 
the system becomes the quasi-Wigner crystal 
caused by the forward scattering for $\beta \leq 1$ and
the system becomes the TL liquid when $\beta > 1$.
I see that
\begin{eqnarray}
\Phi(x) &\equiv& 
\int dx^{'}
\partial_{x} \phi(x,\tau)V(x-x^{'})\partial_{x^{'}} \phi(x^{'},\tau)
\label{Phidef}
\end{eqnarray}
has the scaling dimension $x_{g}=\beta +1$ for $1< \beta < 3$ and
4 for $\beta > 3$. As the weak logarithmic corrections appear
for $\beta$=odd, I here 
distinguish $\Phi(x)$ for $\beta$=3 from
the scaling functions.
I also find the consistency on these scaling
dimensions by CFT.
%For the case $\beta = 1, 3$ I can derive the renormalization 
%group eq. of $g$ and the
%velocity (see Appendixes.).
%
%\subsection*{The finite size scaling of energies}
%It is nontrivial problems if the CFT describes the 
%systems with the long-range interactions. 
%Here I investigate whether the results of the ordinary local CFT 
%are applicable to the Tomonaga-Luttinger liquid with the 
%long-range forward scatterings. 
By using the first order 
perturbation,
I can know the effects of the long-range forward scattering.
Based on CFT,
the finite size scalings of energies for no perturbations are given
\cite{Cardy1,Cardy2,Cardy3,Cardy4} by
\begin{eqnarray}
   \Delta E_{n} &=& \frac{2\pi v x_{n}}{L} \nonumber \\
    E_{g} &=& e_{g}L -\frac{\pi v c}{6L}, \label{scale}
\end{eqnarray} 
where $x_{n}$ is the scaling dimension of the primary field denoted 
by $n$, $v$ is the sound velocity and $c$ is the central charge. 
Considering the LRI,
I can extract the corrections to these 
energy size scalings(see Appendixes.): 
\begin{eqnarray}
\Delta E_{n} &=& \frac{2\pi v x_{n}}{L}(1+
\frac{g(0)}{x_{n}} \frac{{\rm const.}}{L^{\beta-1}}
%+g \frac{1}{L^2}
+O(1/L^2)) \label{scaling}
\nonumber \\
E_{g} &=& (e_{g}+ g(0) {\rm const.})L
-\frac{\pi v }{6L}(c+g(0) \;{\rm const.}+
g(0) \frac{{\rm const.}}{L^{\beta-1}}
%+g \frac{1}{L^2}
+O(1/L^2)),\nonumber \\
\label{fscale}
\end{eqnarray}
where $\beta (> 1)$ is not odd. And
the constants are the functions of $\beta$.
Note that for $\beta=\rm odd$ cases, the 
logarithmic corrections appear. They 
correspond to the integer points of the modified Bessel function, which
appear in the long-wave behaviours (\ref{longwave}).
I can reproduce these anomalies
for $\beta= \rm odd$ by the CFT. 
Moreover from CFT I can show that there 
are the anomalies in the general excitations 
and the ground state energy.
The details are shown in Appendixes.
The $O(1/L^2)$ terms come from the irrelevant field
$L_{-2} \bar{L}_{-2} {\bf 1}$ and the long-range g term. 
The first eq. of (\ref{scaling}) 
means that the long-range forward scattering $\Phi(x)$
%\begin{eqnarray}
%g\int dx^{'}
%\partial_{x} \phi(x,\tau)V(x-x^{'})\partial_{x^{'}} \phi(x^{'},\tau)
%\end{eqnarray}
has the scaling dimension $x_{g}=\beta +1$ for
$1< \beta < 3$ and 4 for $\beta > 3 $ effectively.
These respective scaling dimensions 
are consistent with the estimation from 
the renormalization group eqs.
%(\ref{RG}) and (\ref{RGtype2})
of $g$, that I mentioned above (see Appendixes.).

The energy finite size scalings (\ref{fscale}) mean that
the LRI has the higher order influences than $1/L$
to the excitation energy and the LRI affects the $1/L$ term in the finite size
scaling of the ground state energy. 
Here I note that it becomes 
difficult to calculate the central charge from 
finite size scalings (\ref{scale}) unless the effects 
of the LRI to $O(1/L)$ terms are known.

It is
notable to compare the eqs. (\ref{fscale})
with the case where the perturbations 
are of short-range type. Ludwig and Cardy calculated the 
contributions of the short-range perturbation \cite{Ludwig}.
The results for the irrelevant perturbation, 
$-g \sum_{r} \phi(r)$, which has the scaling dimensions
$x > 2$ are  
\begin{eqnarray}
\Delta E_{n} &=& \frac{2\pi v x_{n}}{L}(1+\frac{g(0)}{x_{n}}
C_{nng}(\frac{2\pi}{L})^{x-2}) \nonumber \\
E_{g} &=& (e_{g}+ g (0) {\rm const.})
 L -\frac{\pi v }{6L}(c+ g(0)^{2} \frac{{\rm const.}}{L^{2x-4}}
+O(1/L^{3x-6}) ),
\end{eqnarray}
where the $O(g)$ terms do not appear in the ground state scaling
because we set 
$\langle \phi \rangle =0$ for the short-range interaction. 
These results mean that the $x > 2$ irrelevant field has influences
of the higher order to 
the finite size scalings (\ref{scale}). 
And their result contains parts not so simple. 
There are the special points of scaling dimension $x=1,3,5,$ and $x=2$
which is related to the appearance of logarithmic corrections.

To the contrary, I see $\langle \Phi \rangle \neq 0$
in the long-range case, where $\Phi$ is defined in eq. (\ref{Phidef}). 
The LRI
gives the $O(1/L)$ intrinsic influence to
the finite size scaling of the ground state energy, as appearing
in the scalings (\ref{fscale}), even if the LRI
is irrelevant, that is, $x_{g} > 2$.
%It is known that the $1/r^{2}$ type Sutherland model exhibits the same
%situations. Now I find the 
%The solution of RG eqs. for $1< \beta <3$ is 
%\begin{eqnarray}
%g(l) &=& g(0) e^{(1-\beta)l}=g(0) e^{(1-\beta)\ln L}
%=g(0)\frac{1}{L^{\beta-1}},
%\end{eqnarray} 
%where I use $l=\ln L$.  
%Finally I obtain the accurate finite size scaling:
%\begin{eqnarray}
%\Delta E_{n} &=& \frac{2\pi v x_{n}}{L}(1+\frac{{\rm const.}}{L^{2(\beta-1)}}
%+\frac{{\rm const.}}{L^{\beta+1}}+O(1/L^2))
%\nonumber \\
%E_{g} &=& e_{g}L -\frac{\pi v c}{6L}(1+\frac{{\rm const.}}{L^{2(\beta-1)}}
%+\frac{{\rm const.}}{L^{\beta+1}}+O(1/L^2)).
%\end{eqnarray} 
%
\section{Numerical calculations}
%I have investigated the properties of the energy scaling and 
%derived the corrections terms caused by the long-range forward scatterings.
Through
the Jordan-Wigner transformation,
I transform the model (\ref{Hamillng}) to $S=1/2$ spin Hamiltonian
for the numerical calculations:
\begin{eqnarray}
H &=& - \sum_{j}(S^{+}_{j} S^{-}_{j+1} + {\rm h.c}) +
\frac{g}{2}
\sum_{i\neq j} S^{z}_{i} V(|i-j|) S^{z}_{j}.\label{spinHamil}
\end{eqnarray} 
I impose the periodic boundary condition ${\bf S}_{L+1}={\bf S}_{1}$
to this model.
%and put $\rho \equiv <n>=1/2$, that
%is $S^{z}_{tot}=0$. 
Using the Lanczos algorithm I perform the numerical 
calculations for the Hamiltonian (\ref{spinHamil}).

I have found analytically 
the corrections to the energy scalings
(\ref{scale}) caused by the long-range forward scattering. 
If the oscillating Umklapp process term of (\ref{actotal})
is irrelevant and does not disturb the energy scalings,
the finite size corrections due to the forward 
scattering are expected to appear in the excited state energies
and the ground state energy. I attempt to
detect the contribution of the forward 
scattering.

\begin{figure}[htbp]
\rotatebox{-90}{\scalebox{0.5}{\includegraphics{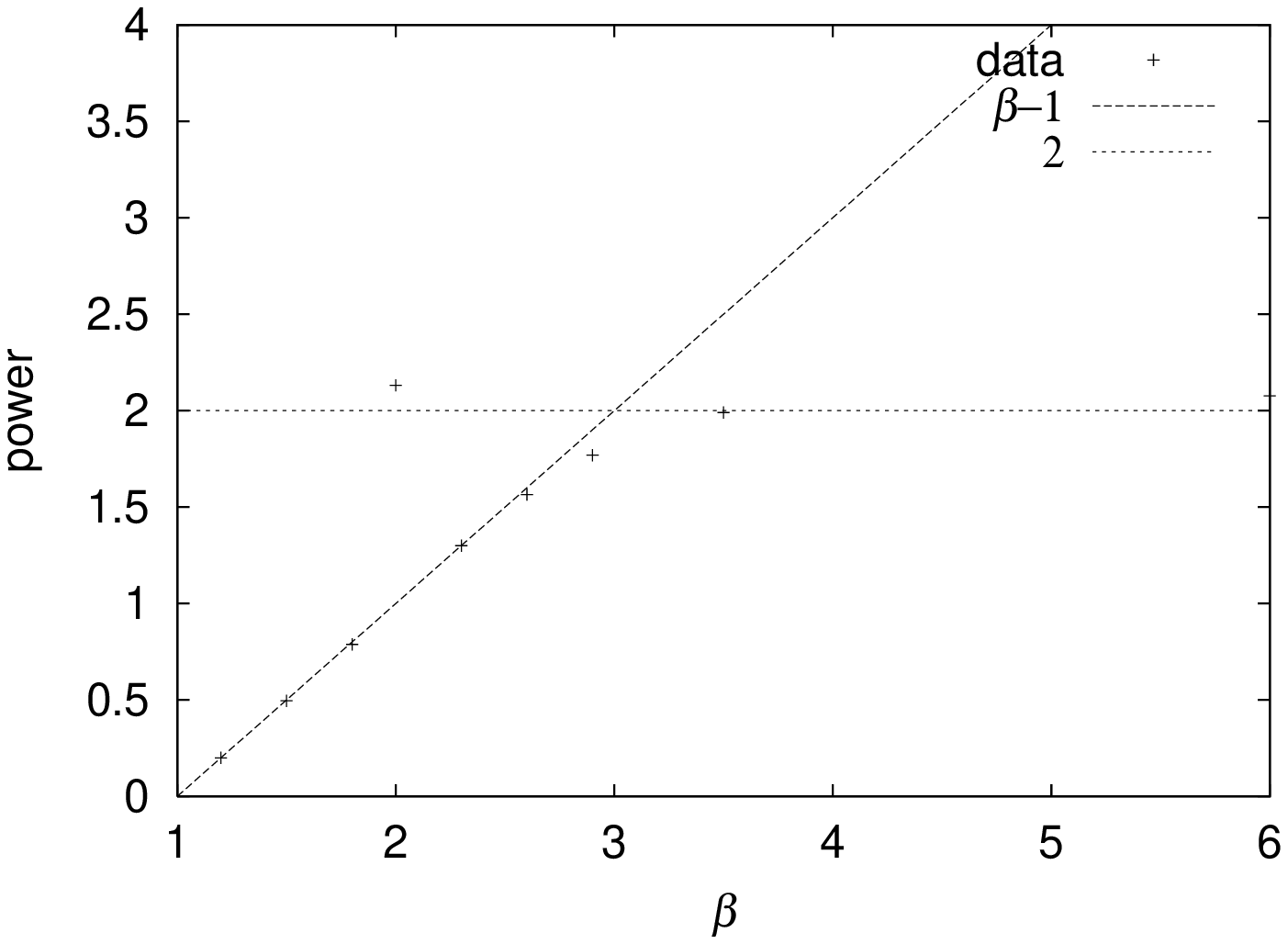}}}
\caption{The numerically calculated
powers $c$ in the excitation energy 
$L \Delta E (m=1/L)$ are shown versus $\beta$ for $g=0.5$.
Here I use the scaling form: 
$L\Delta E = a+ \frac{b}{L^{c}}+\frac{d}{L^{2}}$, where 
$a,b,c$ and $d$ are determined numerically.
If the LRI is not present,
the energy finite size scaling must take the form
:$L\Delta E = A + \frac{B}{L^{2}}+O(1/L^{4})$,
where $A$ and $B$ are constant values.
}
\label{fig4}
\end{figure}

\begin{figure}[htbp]
\rotatebox{-90}{\scalebox{0.5}{\includegraphics{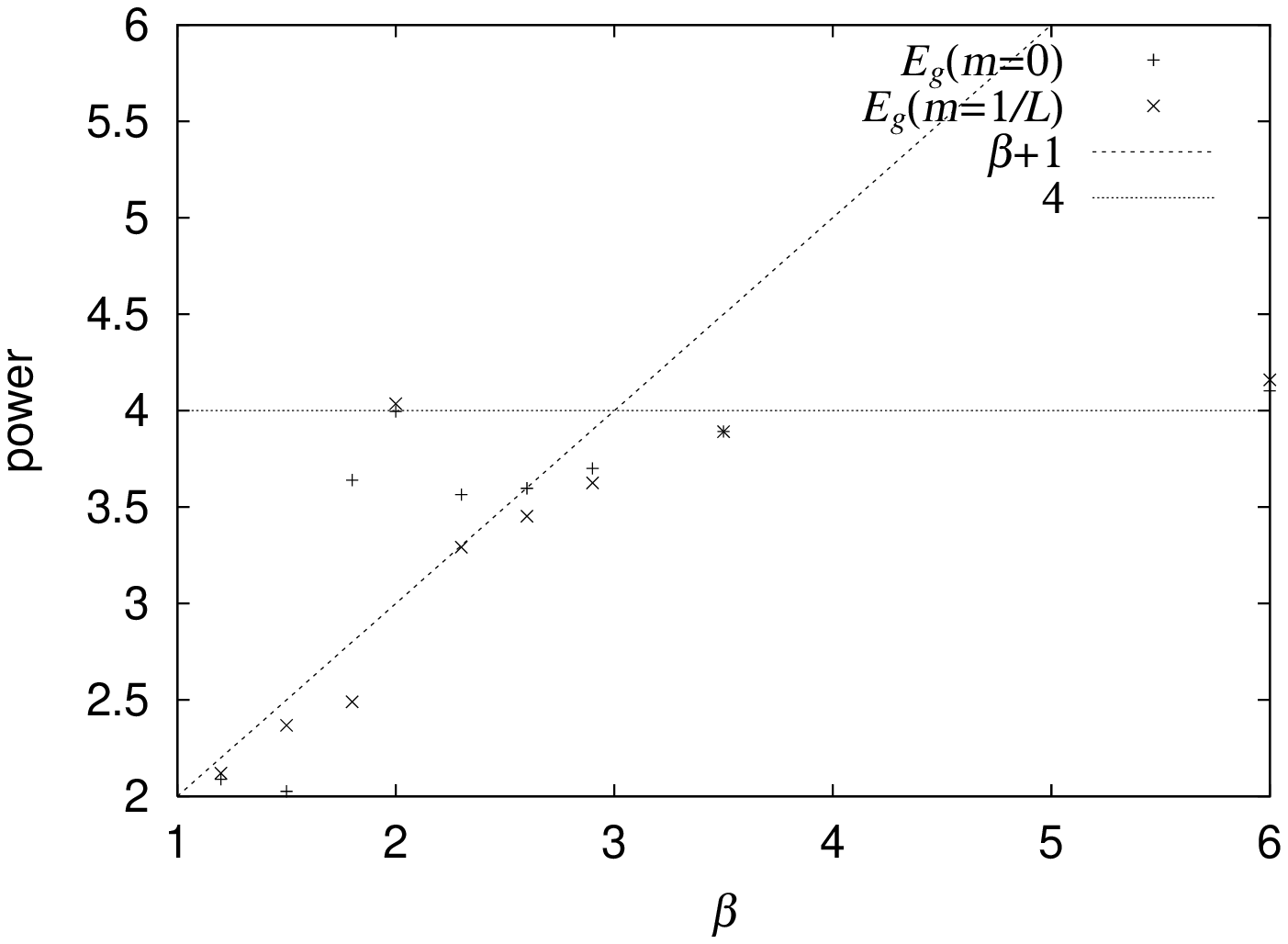}}}
\caption{
The numerically calculated powers $d$ in
the ground state energies
$E_{g}(m=1/L)/L$ and $E_{g}(m=0)/L$ 
are shown versus $\beta$ for $g=0.5$.
Here I use the scaling form:
$E_{g}/L = a+ \frac{b}{L^{2}} + \frac{c}{L^{d}}$, where 
$a,b,c$ and $d$ are determined numerically.
If the LRI is not present,
the energy finite size scaling must take the form
:$E_{g}/L = A + \frac{B}{L^{2}}+\frac{C}{L^{4}}$,
where $A,B$ and $C$ are constant values.
}
\label{fig5}
\end{figure}

%density $\rho \equiv \sum_{j} \rho_{j} /L$.
%% revise
%The size dependences of the excitation energy $\Delta E (m=1/L)$
%and the ground state energy $E_{g}(m=0)$, $E_{g}(m=1/L)$
%for $g=0.5$ are shown in Fig. \ref{data1}, Fig. \ref{data2} and
%Fig. \ref{data3} respectively. 
%Here I define the magnetization $m \equiv \sum_{j} S^{z}_{j} /L$ which 
%is the conserved quantity.
%\begin{figure}[htbp]
%\rotatebox{-90}{\scalebox{0.5}{\includegraphics{exc.ps}}}
%\caption{The size dependences of the excitations 
%$\Delta E (m=1/L)$ for $g=0.5$.}
%\label{data1}
%\end{figure}
%\begin{figure}
%\rotatebox{-90}{\scalebox{0.5}{\includegraphics{eg.ps}}}
%\caption{The size dependences of the ground state energy per site
%$E_{g}(m=0)/L$ for $g=0.5$.}
%\label{data2}
%\end{figure}
%\newpage
%\begin{figure}[htbp]
%\rotatebox{-90}{\scalebox{0.5}{\includegraphics{egsz1.ps}}}
%\caption{The size dependences of the ground state energy per site
%$E_{g}(m=1/L)/L$ for $g=0.5$.}
%\label{data3}
%\end{figure}
%the excitation energy $\Delta E (m=1/L)$
%and the ground state energy $E_{g}(m=1/L)$
%for avoiding the effects from the Umklapp process. 
I numerically calculate the size dependences of the excitation energy $\Delta E (m=1/L)$
and the ground state energy $E_{g}(m=0)$, $E_{g}(m=1/L)$
for $g=0.5$. Here I define the magnetization $m \equiv \sum_{j} S^{z}_{j} /L$ which 
is the conserved quantity.
%is the conserved quantity..
% are shown in Fig. \ref{data1}, Fig. \ref{data2} and
%Fig. \ref{data3} respectively. 
Fitting
the one particle excitation energy as 
$L\Delta E (m=1/L)= a+\frac{b}{L^{c}}+\frac{d}{L^{2}}$, I
show the power $c$ versus the powers $\beta$
in Fig. \ref{fig4}. 
I see the power $c$
agrees with theoretical predictions: $\beta -1$ except for $\beta =2$.
I shall discuss the $\beta =2$ case later.
%The $\beta =2$ case is the solvable. The disturbances may be explained.
Fitting the ground state energy per site
as $E_{g}/L = a+ \frac{b}{L^{2}} + \frac{c}{L^{d}}$, I plot
the powers $d$ versus $\beta$ in Fig. \ref{fig5}.
I see that 
the power $d$ do not show 
agreements with theoretical predictions $\beta +1$ 
in $E_{g}(m=0)/L$. 
These disagreements may be caused by the oscillating
Umklapp process which becomes relevant at only $m=0$
filling. 
On the contrary,
the oscillating Umklapp process is irrelevant at $m \neq 0$. 
Actually, 
in Fig. \ref{fig5},
I see that 
the power $d$ show
agreements with theoretical predictions $\beta +1$
in $E_{g}(m=1/L)/L$ 
except for $\beta =2$.
%I shall discuss $\beta =2$ case just later.

\begin{figure}[htbp]
\rotatebox{-90}{\scalebox{0.5}{\includegraphics{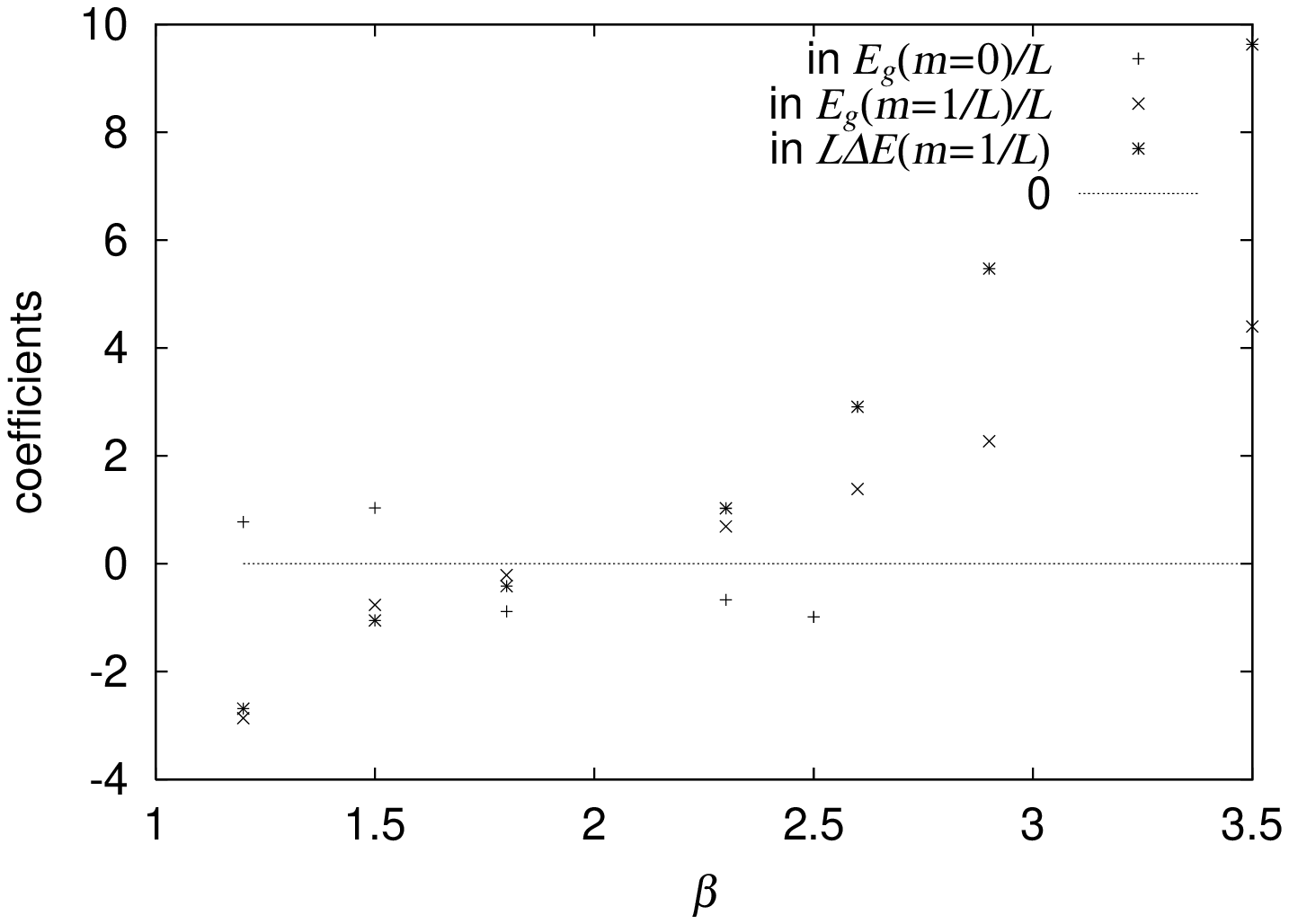}}}
\caption{
I show the numerically obtained
coefficients of $1/L^{d}$
in the size scalings
$E_{g}(m=0)/L,E_{g}(m=1/L)/L = a+ \frac{b}{L^{2}} + \frac{c}{L^{d}}$
and the numerically obtained coefficient of $1/L^{c}$
in the size scaling
$L\Delta E (m=1/L) = a+ \frac{b}{L^{c}}+\frac{d}{L^{2}}$.
%Here I use the data shown in
%Figs. \ref{data1},\ref{data2} and \ref{data3}.  
I observe that 
the coefficient of 
$1/L^{d}$ in $E_{g}(m=1/L)/L$ and
the coefficient of 
$1/L^{c}$ in $L \Delta E(m=1/L)$ 
become small around $\beta =2$.
The coefficients of $1/L^{d}$
in $E_{g}(m=0)/L$
show the different behaviour
from that in $E_{g}(m=1/L)/L$.
This difference may be caused by the spatially oscillating Umklapp process term.}
\label{coeffbeta}
\end{figure}

As I stated above, for $\beta =2$,
the power $c$ in the excitation energy
$L\Delta E(m=1/L)
= a+ \frac{b}{L^{c}}+\frac{d}{L^{2}}$ apparently shows disagreement with theoretical 
value $\beta -1$ and likewise for $\beta =2$, the power $d$
in the ground state energy 
$E_{g}(m=1/L)/L
= a+ \frac{b}{L^{2}} + \frac{c}{L^{d}}$ apparently shows
disagreement with theoretical value $\beta +1$. 
I investigate the reason for these disagreements.
In Fig. \ref{coeffbeta} I show the numerically obtained
coefficient of $1/L^{d}$ 
in the size scalings
$E_{g}(m=0)/L,E_{g}(m=1/L)/L = a+ \frac{b}{L^{2}} + \frac{c}{L^{d}}$
and the numerically obtained coefficient of $1/L^{c}$
in the size scaling
$L\Delta E (m=1/L) = a+ \frac{b}{L^{c}}+\frac{d}{L^{2}}$.
I observe that
the coefficient of $1/L^{c}$ in $L\Delta E(m=1/L)$
and
the coefficient of $1/L^{d}$ in $E_{g}(m=1/L)/L$
become small around $\beta =2$.
So for $\beta =2$, 
$1/L^{2}$ dependence appears rather than
$1/L$ in $L\Delta E(m=1/L)$ (see Fig. \ref{fig4}.).
Likely for $\beta =2$, 
$1/L^{4}$ dependence appears 
rather than $1/L^{3}$ in $E_{g}(m=1/L)/L$ (see Fig. \ref{fig5}.).
%Around $\beta =2$ the coefficients of the $1/L^{\beta +1}$ correction
%become small, as appearing in Fig. \ref{coeffbeta}. 
%So for $\beta =2$ the $1/L^{4}$ dependences appear
%rather than $1/L^{\beta + 1}$ in Fig. \ref{fig5}. 
I observe that the coefficient of $1/L^{d}$ in $E_{g}(m=0)/L$ show
the different behaviour from those in $E_{g}(m=1/L)/L$
in Fig. \ref{coeffbeta}.
This difference may come from
the spatially oscillating Umklapp process that opens the gap at $m=0$ and 
disturbs the finite size scaling.    

\begin{figure}[htbp]
\hspace{0.51cm}
\rotatebox{0}{\scalebox{1.0}{\includegraphics{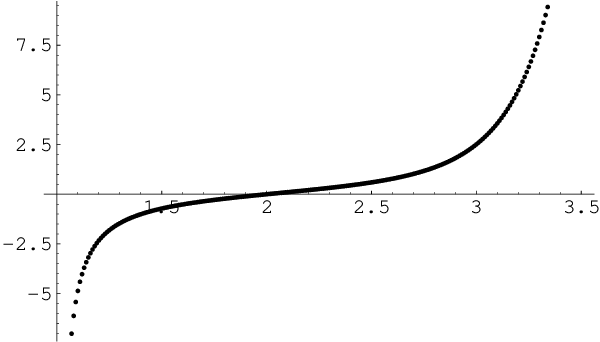}}}
%\caption{(a)}
%\label{integral1}
%\end{figure}
%\begin{figure}[htbp]

\vspace{1cm}
\hspace{1cm}
\rotatebox{0}{\scalebox{0.6}{\includegraphics{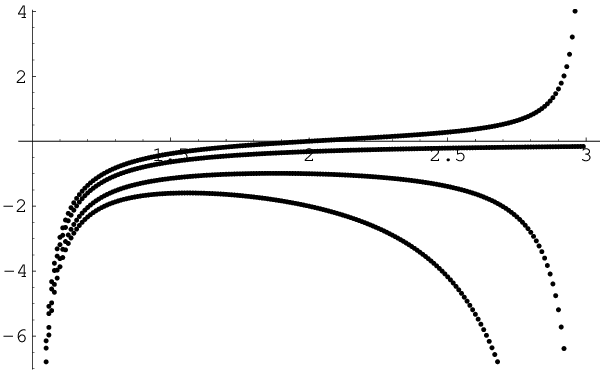}}}
%\caption{(b)}
\caption{(a) $A(\beta)$, the coefficient of $1/L^{\beta}$, in the eq. (\ref{appscale}) is 
shown. I see that $A(\beta)$ has zero point close to $\beta =2$.
This curve coincides with the results
from the numerical calculation in the tight-binding model shown in Fig. \ref{coeffbeta}.
(b) $A(\beta)$, the coefficient of $1/L^{\beta}$, in the eq. (\ref{fscaleapp}) is 
shown for some $s$. Analytically only $s=0$ is meaningful for particle excitations.
$A(\beta)$ for $s=0$ has zero point close to
$\beta =2$. This coincides with the results from  
the numerical calculation in the tight-binding model shown in Fig. \ref{coeffbeta}.}
\label{integral2}
\end{figure}

I can obtain $A(\beta)$ in the scalings (\ref{appscale}) and
(\ref{fscaleapp}) by evaluating the integrals. The results are shown
in Fig. \ref{integral2} (a) and (b).
The analytical $A(\beta)$ in the scalings (\ref{appscale}) and
$A(\beta,s)$ in the scalings (\ref{fscaleapp}) for $s=0$ fit 
with the points in Fig. \ref{coeffbeta} well.
The curve only for $s=0$ in Fig. \ref{integral2}(b) shows the good fitting.
This point shall be discussed later.
These reveal that
the present numerical calculation of the tight-binding model agrees with the CFT analysis of the
long-range forward scattering. 
 
%
%I would transfer 
%the problems why the coefficients of the energy scalings become small
%for $\beta =2$ to future works.
%
%That may be explained by using Bethe ansatz solutions
%and the obtained scalings shown in Appendixes.
% 
%The oscillating terms may disturb the TLL and
%cause the mass gap. 
%I numerically investigate how long range the TLL phase survive for
%the strength of the long-range interactions.
%By making use of eqs. (2.15), I can calculate the
%compressibility $\chi=K/v$ and the drude Iight $D=vK$ within
%the TLL framework. 
%
%Next
%I numerically show that the tight-binding model 
%(\ref{Hamillng}) can be described by the $c=1$ CFT
%\footnote{
%It is not still trivial that
%the tight-binding model with the long-range interactions 
%$1/r^{\beta}$ is
%described by the CFT. }.

\begin{figure}[htbp]
 \rotatebox{-90}{\scalebox{0.5}{\includegraphics{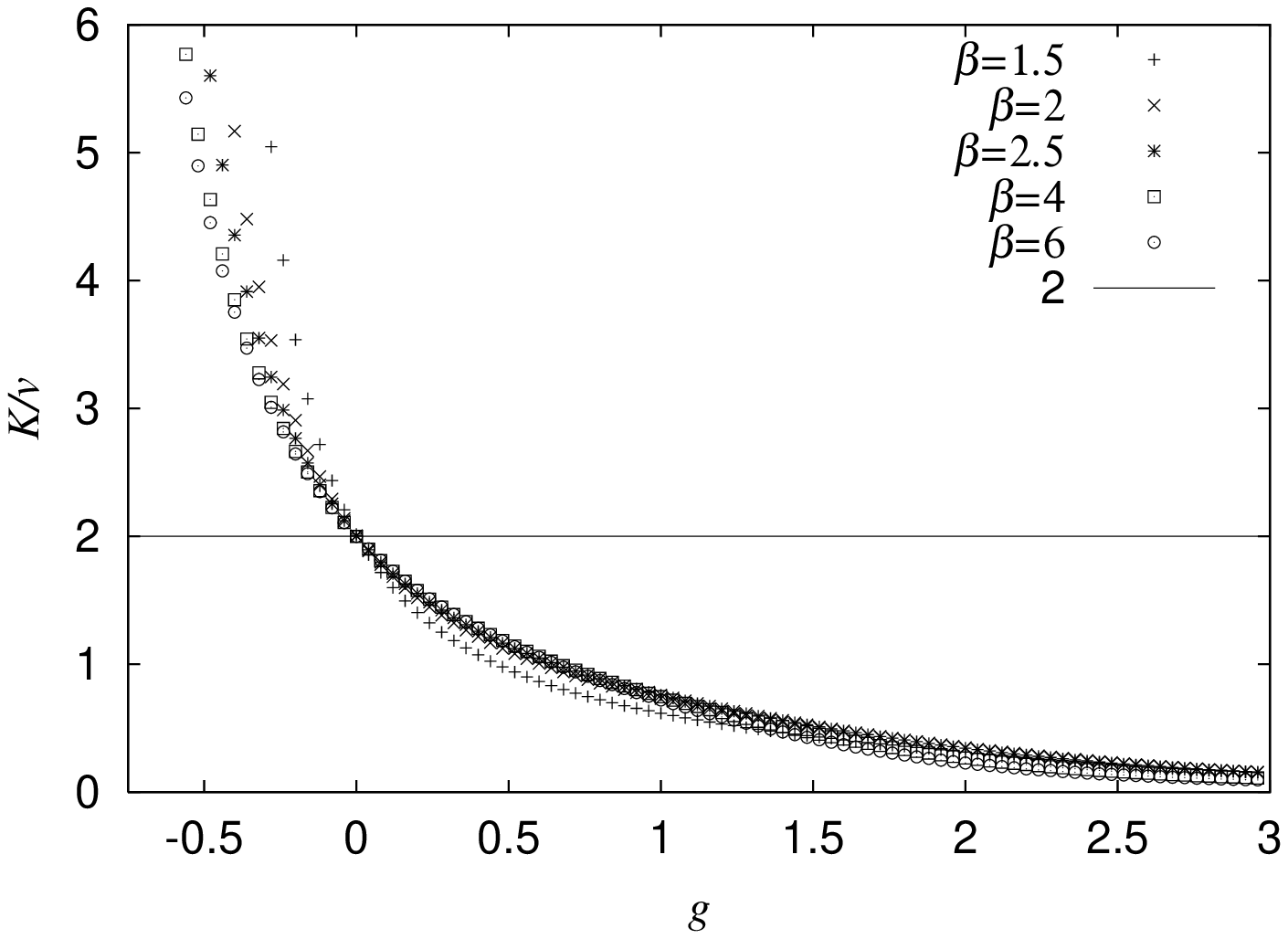}}}
\caption{
The extrapolated $K/v (=\chi /2) $ 
is plotted versus the strength $g$.
I use the scaling form $v/K(L)=v/K(\infty)
+\frac{a}{L^{\beta -1}}$ for $\beta < 3$,
%+\frac{b}{L^{2}}$
and $v/K(L)=v/K(\infty)
+\frac{a}{L^{2}}$
%+\frac{b}{L^{4}}$ 
for $\beta \geq 3$,
where $v/K(\infty)$, $a$ is
%and $b$ are 
determined
numerically. }
\label{fig6}
\end{figure}

\begin{figure}[htbp]
\rotatebox{-90}{\scalebox{0.5}{\includegraphics{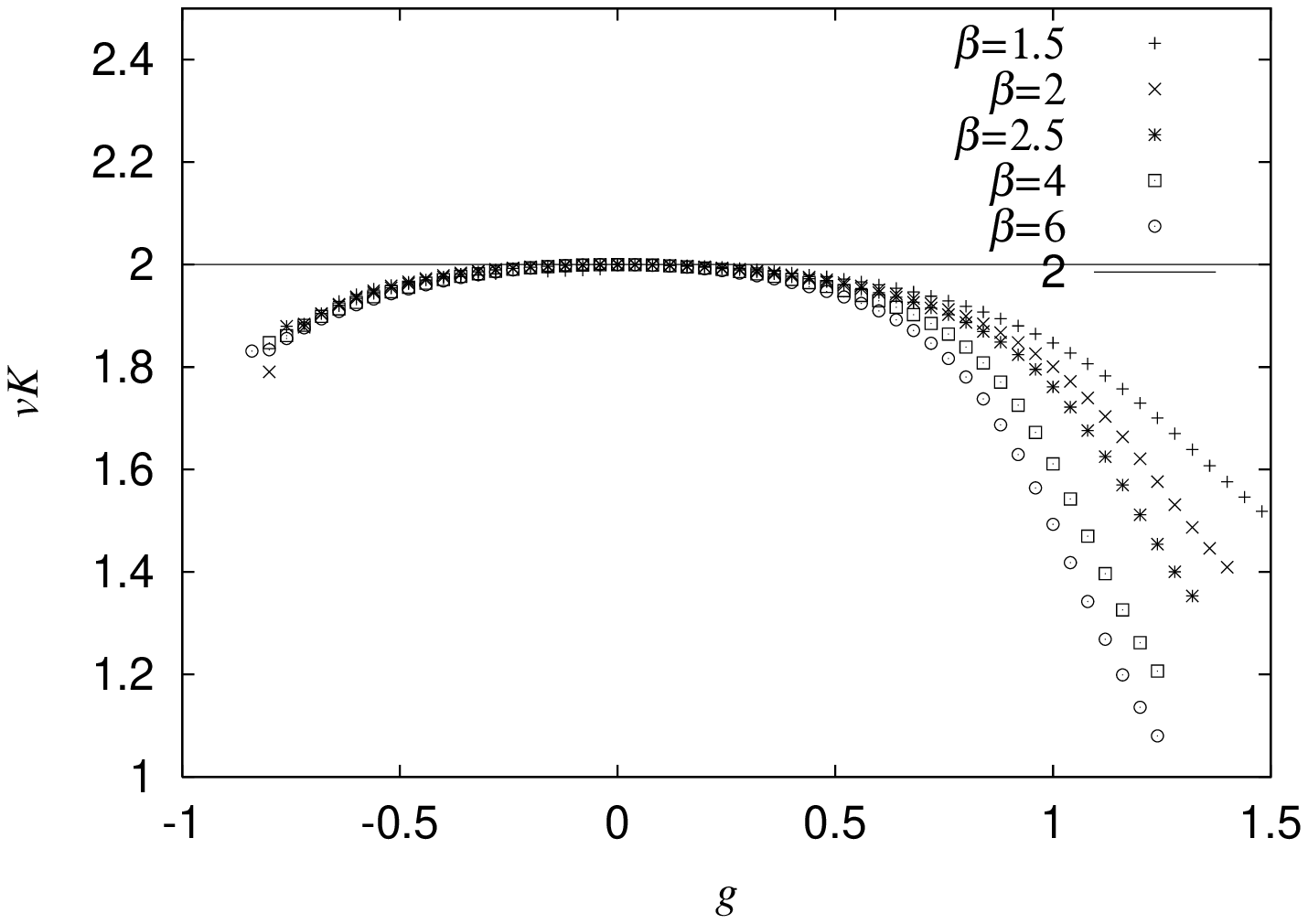}}}
\caption{The extrapolated $vK(=2D)$ is plotted versus the strength
 $g$. I use the scaling form $L\Delta E = a+ \frac{b}{L^{c}}$,
where a,b and c are determined numerically.
}
% I use the same scaling as compressibility.}
\label{fig7}
\end{figure}

Next I survey whether the long-range tight-binding model satisfies the necessary condition of CFT.
The operator $\cos {\sqrt 2} \phi$ 
has the scaling dimensions $K/2$ and 
the operator $e^{\pm i \sqrt 2 \theta}$ has
$1/2K$ in the regime of  
the TL liquid. The two quantities $2K/v$
and $vK/2$ are the compressibility and the Drude weight
respectively in the regime of the TL liquid. If $c=1$ CFT is valid to
the tight-binding model with the LRI, 
the two quantities are related to the two excitations
with the symmetries $q=\pi,m=0$ and $q=\pi,m=1/L$ respectively:
\begin{eqnarray}
 2K/v &=& 1/(L\Delta E(m=1/L, q=\pi)) \equiv \chi \nonumber \\
 vK/2 &=& L\Delta E(q=\pi) \equiv D. \label{fromCFT}
\end{eqnarray}
I show the numerically calculated quantities 
%$1/(L\Delta E(S_z=1, q=\pi))=
$\chi$
and 
%$L\Delta E(q=\pi)= 
$D$ in Fig. \ref{fig6} and \ref{fig7},
where I use the sizes $L=16,18$ and $20$ and
extrapolate the data. 
For $g < 0$, $\chi$
(which is the susceptibility, irrespective of the CFT arguments) 
exhibits the rapid increase which suggests the phase 
separation. In spin variables' language for (\ref{Hamillng}),
this phase separation is nothing but the ferromagnetic phase. 
Hence for the larger $\beta$ the 
point of the phase separation approaches to $-1$. 
For $g > 0$ I see the weak tendency
that the the quantity $\chi$
becomes smaller as $\beta$ is smaller for $g$ less than about 1.   
I find that the the quantity $D$ of $g > 0$
become larger as $\beta$ approaches to $\beta=1$.

\begin{figure}[htbp]
\rotatebox{0}{\scalebox{0.6}{\includegraphics{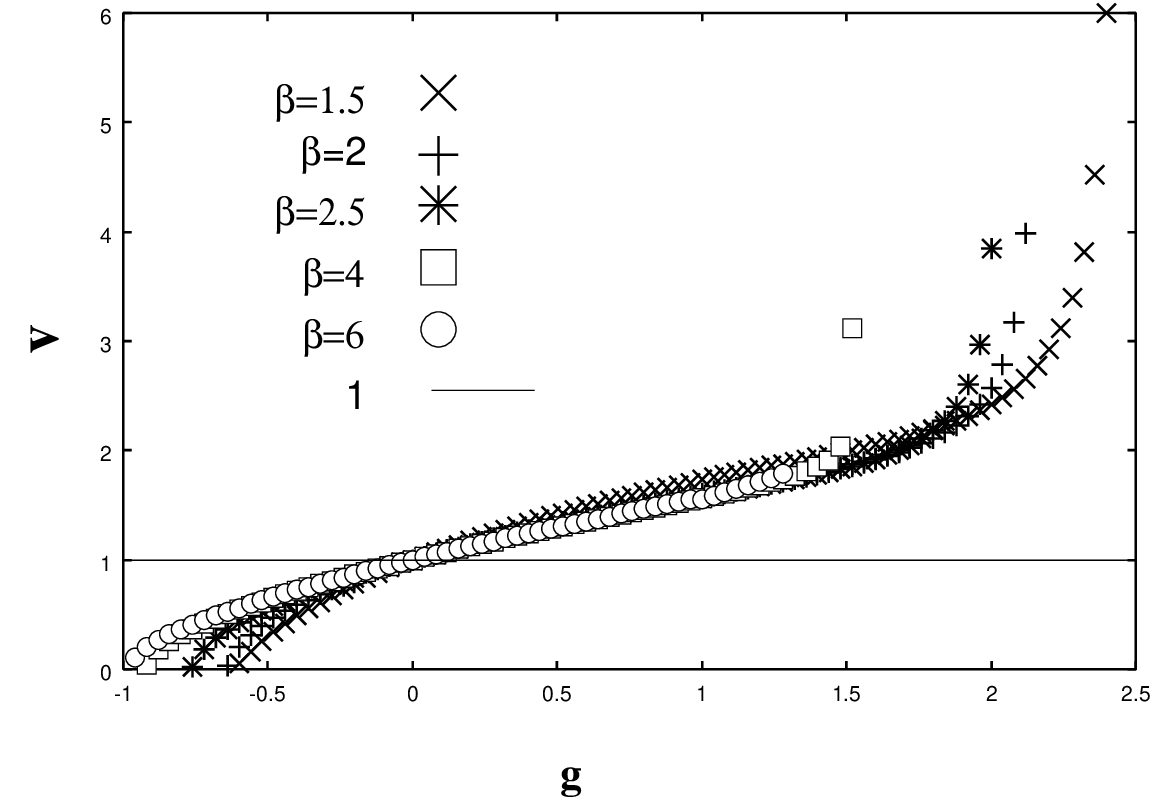}}}
\caption{The extrapolated 
spin wave velocity $v$ is plotted versus the strength $g$.
I use the scaling form $L\Delta E = a+ \frac{b}{L^{c}}$,
where a,b and c are determined numerically.}
%I use same scaling as compressibility.}
\label{fig8}
\end{figure}

In Fig. \ref{fig8} I plot the velocity versus
the strength $g$ for the various powers $\beta$, where
the velocity is defined by
\begin{eqnarray}
 v&=&\frac{L}{2\pi}\Delta E(q=2\pi/L).
\end{eqnarray}
I see that the velocities are finite values for $\beta >1$ 
, as is expected.
%\footnote{In the Coulomb interactions 
%case $\beta=1$ the velocity show the Iak divergence (see Appendixes.).}. 
There are the points where the velocities 
are zero, implying the phase separation.

\begin{figure}[htbp]
\rotatebox{-90}{\scalebox{0.5}{\includegraphics{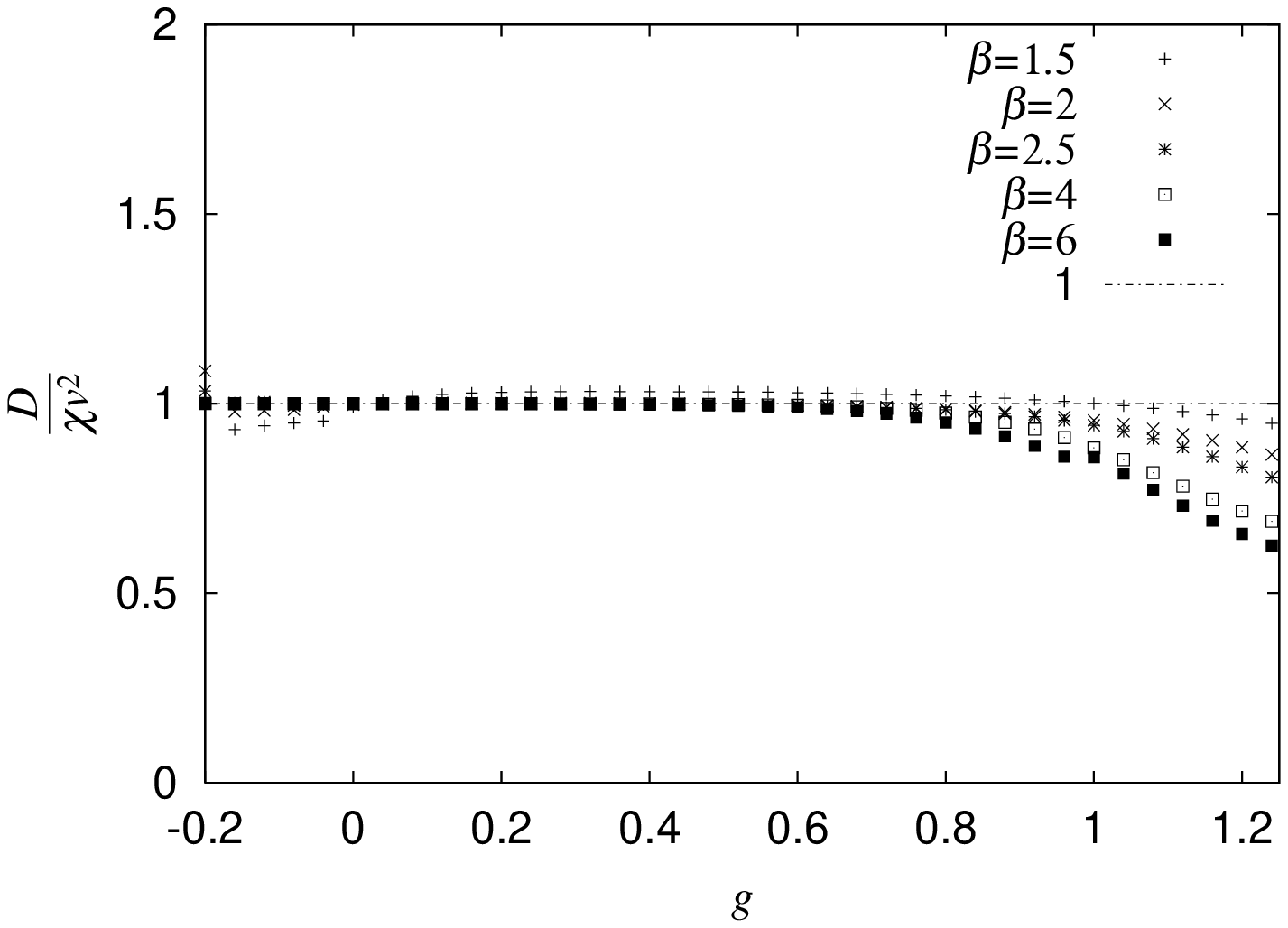}}}
\caption{The normalization $ \frac{D}{\chi v^2}$ is plotted versus the
 strength $g$.}
\label{fignorm}
\end{figure}

In Fig. \ref{fignorm} I plot the quantity $\frac{D}{\chi v^2}$ versus
the strength $g$ for the various powers $\beta$.
%, where
%$D$, $\chi$ and $v$ are defined in \ref{} respectively.
If the present system is described by $c=1$ CFT,
this quantity is 1 from eqs. (\ref{fromCFT}). I find the regions where
$\frac{D}{\chi v^2}=1$ in Fig. \ref{fignorm}.
The regions become wider as $\beta$ 
approaches to $1$ for $g > 0$. For larger $g$, the normalization
breaks owing to the generations of mass.
%I shall discuss this point later. 
%Reversely for $g< 0$ the region is smaller as $\beta$ goes to $1$.
%

%I calculate
%the effective central charge from the 
%ground state energy scalings
%(\ref{scale}), where I use the numerical data of
%$L=16,18$ and $20$.
%The obtained central charges 
%are plotted versus $g$ in Fig. \ref{fig10}.
%\begin{figure}[htbp]
%\rotatebox{0}{\scalebox{0.6}{\includegraphics{cent.eps}}}
%\caption{The effective central charges for various $\beta$
%are plotted versus $g$. I use
%the scaling form 
%$E_{g}/L = a+ \frac{b}{L^{2}}$, where a and b are determined numerically.
%From the coefficient $b$ and the velocity I calculate the effective 
%central charge.}
%\label{fig10}
%\end{figure}
%In Fig. \ref{fig10}, 
%I observe that the effective central charge deviates from 1
%and the deviation is more remarkable as $\beta$ becomes larger.
%As there is the deviation from $c=1$ due to the
%long-range forward scatterings in the finite size
%scalings (\ref{fscale}),
%the deviation of the effective central charge from 1 agrees with the
%theoretical predictions.
%But it seems strange apparently that 
%In the later section,
%I shall discuss these behaviours.
%the deviations from 1
%are more remarkable as $\beta \rightarrow$ larger
\section{Discussion}
I have investigated 
the system with the $1/r^{\beta}$ interaction
by applying CFT to it and by the numerical calculation. At first I have analyzed
TL liquid with the $1/r^{\beta}$ forward scattering by 
utilizing the CFT and I have found
that the $1/r^{\beta}$ forward scattering works as higher order 
corrections in the excitation energy, whereas the effective 
central charge in the scaling of 
the ground state energy depends on the interaction and it deviates from $c=1$. 
The deviation 
are like the solvable $1/r^{2}$ models \cite{Haldane2,Sutherland,Sutherland2,Kawakami2}.
Next I have numerically calculated the ground state energy 
and excitations energies in the tight-binding model with 
$1/r^{\beta}$ interaction, which is expected to include
the above $1/r^{\beta}$ forward scattering in the low
energy. The numerical results are in accordance 
with the analysis with CFT of the long-range forward scattering. 
Furthermore I have numerically checked the normalization $\frac{D}{\chi v^2}=1$, which 
is the necessary condition for $c=1$ CFT.

For $\beta \approx 2$, the coefficient $A(\beta)$ in the ground state energy 
vanishes. This seems to correspond to the exact solution for 
$\beta=2$ \cite{Kawakami2} which states that
the finite size scaling of ground state has no higher order term than $1/L$. 
The coefficient $D(\beta)$ of $1/L^{3}$ in eq. (\ref{Db}) does not vanish for $\beta=2$.
However the present argument is the first order perturbation theory. With higher order treatments,
I may clarify this. In any case, with consistency in many points
I could construct CFT in the system with non-local interaction. 
  %except for the $\beta =2$ case. 
%Though
%I can not refer to the disagreement of the $\beta =2$ case 
%unfortunately here, this problem may be explained by the exact solutions
%in Ref. \cite{Kawakami2}.

The numerical calculations in the tight-binding model support
the finite size scalings (\ref{appscale}) and (\ref{fscaleapp}).
In one particle excitation energy $L \Delta E(m=1/L)$, the coefficients of 
$1/L^{\beta}$ fit with $s=0$ case in Fig. \ref{integral2}. The 
coefficients from the long-range forward scattering
are related with the operator product expansion.
I can prove that only $s=0$ case
is relevant for the particle excitation.
Using $<\varphi(z) \varphi(z^{'})>= -\frac{K}{4}\ln (z-z^{'})$ and
$<\bar{\varphi}(\bar{z}) \bar{\varphi}(\bar{z^{'}})>=
 -\frac{K}{4}\ln (\bar{z}-\bar{z^{'}})$,
I confirm the operator product expansions:
\begin{eqnarray}
 \partial \varphi (z) :e^{i \sqrt{2} \theta(z,\bar{z^{'}})}: &=& 
\frac{-i \sqrt{2}}{4} \frac{1}{z-z^{'}} 
:e^{i \sqrt{2} \theta(z^{'},\bar{z^{'}})}: + \rm{ reg.} \nonumber \\
\bar{\partial} \bar{\varphi} (\bar{z}) :e^{i \sqrt{2} \theta(z^{'},\bar{z^{'}})}: &=& 
\frac{i \sqrt{2}}{4} \frac{1}{\bar{z}-\bar{z^{'}}} :e^{i \sqrt{2} \theta(z^{'},\bar{z^{'}})}:
 + \rm{ reg.}
\\
T(z)  :e^{i \sqrt{2} \theta(z^{'},\bar{z^{'}})}: &=& \frac{1}{4K} 
\frac{1}{(z-z^{'})^2} :e^{i \sqrt{2} \theta(z^{'},\bar{z^{'}})}:
+\frac{i \sqrt{2}}{K} \frac{1}{z-z^{'}}
:\partial \varphi(z) (:e^{i \sqrt{2} \theta ( z^{'},\bar{z^{'}})}:) :+ \rm{ reg.}\nonumber \\
\bar{T}(\bar{z})  
:e^{i \sqrt{2} \theta (  z^{'},\bar{z^{'}}  )  }: 
&=& \frac{1}{4K} 
\frac{1}{(\bar{z}-\bar{z^{'}})^2} :e^{i \sqrt{2} \theta(z^{'},\bar{z^{'}})}:
-\frac{i \sqrt{2}}{K} \frac{1}{\bar{z}-\bar{z^{'}}}
:\bar{\partial} \bar{\varphi} (\bar{z}) (:e^{i \sqrt{2} \theta ( z^{'},\bar{z^{'}})}:) :
+ \rm{ reg.}, \nonumber
\label{OPE}
%+O(1/L^2)+O(1/L)) 
\end{eqnarray}
where I define $ T(z)\equiv -\frac{2}{K} (\partial \varphi (z))^2$, 
$ \bar{T}(\bar{z}) \equiv -\frac{2}{K} (\bar{\partial} \bar{\varphi} (\bar{z}))^2$ and
$\theta(z,\bar{z}) \equiv \frac{1}{K}(\varphi (z)-\bar{\varphi} (\bar{z}))$.
From the first and the second eqs., I see $C_{\alpha 1 0}=-i \sqrt{2}/4$,
$C_{\alpha 1 \bar{0}}=i \sqrt{2}/4$
for $\alpha=1$ and 
$C_{\alpha 1 0}=C_{\alpha 1 \bar{0}}=0$ otherwise, where
0( $\bar{0}$ ) and 1 denote 
$\partial \varphi (z)$ ( $\bar{\partial} \bar{\varphi} (\bar{z}))$ ) and 
$:e^{i \sqrt{2} \theta(z,\bar{z})}:$. From the third and the fourth eqs., I see
$:e^{i \sqrt{2} \theta(z,\bar{z})}:$ have the conformal dimension $(1/4K,1/4K)$
and spin 0. As $i(\partial \varphi (z)-\bar{\partial} \bar{\varphi} (\bar{z}))/2$ is 
associated with $\partial_{\sigma}(\sigma)$ for $z=\exp(\frac{2\pi w}{L})$, I obtain
\begin{equation}
<\alpha|\partial_{\sigma}(\sigma)|1> =
\begin{cases} 
\frac{2\pi}{L} i(C_{1 1 0}-
C_{1 1 \bar{0}})/2=\frac{\sqrt{2}}{4} \frac{2\pi}{L} &\text{\rm for $\alpha=1$} \\
0 &\text{\rm otherwise,}
\end{cases}
\end{equation}
which means that only $s=0$ is relevant for the particle excitation and 
the last eq. in (\ref{excLdep}) has no cosine term. 

I would discuss the size effects for $\beta=1$.
As seen in the eqs. (\ref{vKtendency}) and (\ref{fscaleapp}),
the velocity shows the weak divergence for the size and the Luttinger parameter
vanishes gradually for increasing size. This is consistent with the numerical tendency
(see Figs. 8 and 9 in Ref. \cite{Capponi}). 
The size effect of the Drude weight 
(proportional to the charge stiffness) is now given by $v k/2 \sim \rm const.$
as the logarithmic contributions cancel.
The numerical data (see Fig. 7 in Ref. \cite{Capponi}) shows the metallic
behaviour at small and intermediate magnitude interaction strength (larger 
than the CDW transition point $V=2$ by the short range interaction). 
I think that the long-range forward scattering enhances the metallic character.
For fairy large interaction the long-range Umklapp scattering becomes relevant
and the charge stiffness is suppressed.
%About the charge stiffness I mention our numerical calculation in Fig. 6,
%which shows that the Drude Iight
%increases as $\beta \rightarrow 1$. This tendency may correspond to that of 
%Fig. 2 in Ref. \cite{Capponi}. Compared with the short range ($r=1$ in the word of the Ref.),
%the long range ($r>1$) interaction enhances
%the charge stiffness. 
Finally I would add the size effect of the compressibility:
\begin{eqnarray}
L \Delta E (\rho =1/2+1/L) &\equiv& 1/\chi = O(\ln L) \rightarrow \infty, 
\label{chargetend}
%%+O(1/L^2)+O(1/L)) 
\end{eqnarray}
which comes from the results 
(\ref{vKtendency}) by the RG analysis and the CFT arguments.
The compressibility $\chi$ goes to 0 weakly for increasing size.
%These mean the system approaches to
%the crystal (quasi-Wigner crystal), not the TL liquid. 

%The eq. (\ref{chargetend}) 
%may explain the small residue of extrapolated value
%in the charge gap behaviors \cite{Gia}.
%According to these results, 
%the g terms for $\beta =1$ 
%break the critical (TL liquid) 
%behaviors $L\Delta E= {\rm const.}$ for 
%$L \rightarrow \infty$. 
%There is the 
%numerical report \cite{Gia} about the same situations
%appearing in the velocity\cite{Gia}, which shows that the   
%TL liquid is broken by the Coulomb interactions.

%I observed that the deviation from 1 of the effective
%central charge becomes more remarkable
%as $\beta$ goes to a larger value. 
%I believe that this behaviour is due to the following reasons.
%I need the larger size to obtain the more accurate value of the effective
%central charge
%from the finite size scaling of the ground state energy, because the
%size dependences of interactions $V(i-j)$
%become more prominent as $\beta$ is larger.
%The other reason is that
%there are
%the effects from the Umklapp process. 
%As the main reason which I believe,
%the coefficient $C(\beta)$ 
%of $O(1/L)$ term in the energy scalings (\ref{appscale})
%increases as $\beta $ become larger if I choose $\alpha_{0} < 1$. 
%So the deviation from 1 of the effective central charge
%increases as $\beta \rightarrow \rm \; larger$. 
%Such the features may
%be caused by the oscillating terms of the action (\ref{actotal}), which
%can no\
%t
%be explained by from eqs. (\ref{fscale}) unfortunately.

To summarize, within the perturbation theory I have constructed CFT 
in the TL liquid with $1/r^{\beta}$ long-range forward scattering.
I have found that 
the interaction gives the nontrivial behaviour for
$\beta=$odd and $\beta \approx 2$. 
I have numerically checked the finite size scalings obtained 
from CFT in
the tight-binding model with $1/r^{\beta}$ LRI.
Our analysis and numerical calculations exhibit consistency with each other.
%Consequently I predict that the range of TL liquid and the Drude Iight
%increase as the interactions' poIr $\beta$ approaches to 1 which 
%is the Coulomb interaction case. 
%
%
%I thank K. Nomura and A. kitazawa for their comments and encouragements.
\section*{Appendixes}
\section*{1. Renormalization group equation}
\subsection*{$0 < \beta < 1 $ or $1< \beta < 3 $}
I derive the renormalization group equations heuristically.
Let us start from the action (\ref{action2}): 
\begin{eqnarray}
 S &=& \sum_{w} \sum_{q=-\Lambda}^{\Lambda} \frac{2\pi}{K}
(q^2+w^2)|\phi(q,w)|^2+g\sum_{w} \sum_{q=-\Lambda}^{\Lambda}q^2 
V(q)|\phi(q,w)|^2 \nonumber \\ 
   &=& \sum_{w}\{\sum_{q=-\Lambda/b}^{\Lambda/b}
+\sum_{q=-\Lambda}^{-\Lambda/b}+\sum_{q=\Lambda/b}^{\Lambda}\}+g
\sum_{w}\{\sum_{q=-\Lambda/b}^{\Lambda/b}
+\sum_{q=-\Lambda}^{-\Lambda/b}+\sum_{q=\Lambda/b}^{\Lambda}\}.
\label{wavespaction}
\end{eqnarray}
%where $A$ is the constant appearing in the behaviours (\ref{longwave}) and 
%the TL parameter $K^{'}$ is 
The partition function is
\begin{eqnarray}
Z &=& \int {\cal D} \phi_{slow} {\cal D}\phi_{fast}\exp (-
S^{0}_{slow}
-S^{0}_{fast}-S^{g}_{slow}-S^{g}_{fast}). 
\end{eqnarray}
Thus I can integrate out $S_{fast}$ 
($|q|>\Lambda/b$ component) simply and obtain 
\begin{eqnarray}
Z &=& \int {\cal D} \phi_{slow} \exp (-S^{0}_{slow}-S^{g}_{slow}).
\end{eqnarray} 
The remaining procedure of the renormalization is the scale
transformation
\begin{eqnarray}
q \rightarrow q/b, \;w\rightarrow w/b \;\;{\rm and}\;\;
\phi \rightarrow \phi b^2, \label{sctrans}
\end{eqnarray}
where I choose the dynamical exponent 1. The results are 
\begin{eqnarray}
S^{0}_{slow} &\rightarrow & S^{0}\nonumber \\
S^{g}_{slow} &\rightarrow & g\sum_{w}\sum_{q=-\Lambda}^{-\Lambda}
q^2V(q/b)|\phi(q,w)|^2 \nonumber \\
&\rightarrow & gb^{1-\beta}\sum_{w}\sum_{q=-\Lambda}^{-\Lambda}
q^2V(q)|\phi(q,w)|^2,
\end{eqnarray}
where I use $V(q)-A \sim q^{\beta-1}$ 
from the behaviours 
(\ref{longwave}). Hence I obtain the renormalization group eq.
\begin{eqnarray}
\frac{dg(b)}{d b} = (1-\beta)\frac{g(b)}{b}. \label{RGlngrng}
\end{eqnarray}
Substituting $l=\ln b$ into this, I 
obtain renormalization group eqs. 
\begin{eqnarray}
 \frac{dg}{dl}&=& (1-\beta)g \nonumber \\
 \frac{d}{dl}(\frac{v}{K})&=& 0\nonumber \\
 \frac{d}{dl}(\frac{1}{vK})&=&0.
\end{eqnarray}
The TL
parameter K is not renormalized but it shifts due to 
the constant $A$.
\subsection*{$\beta =1$}
The dispersion relation of 
the Coulomb interaction includes the marginal part $w \sim q$ and 
$w \sim q \sqrt{|\ln q|}$ as well as $\beta >1$ case. 
%HoIver 
%it is difficult to know the explicit separated function like 
%eqs. (2.7) and (2.8). Therefore I renormalize by using the bare 
%$V(q) \sim -A \log q + B$.
%
Integrating out the fast moving part, I obtain the effective action of the
slow part
\begin{eqnarray}
 S_{\rm slow} &=& \sum_{w} \sum_{q=-\Lambda/b}^{\Lambda/b} \frac{2\pi}{K}
(v q^2+w^2/v)|\phi(q,w)|^2+g\sum_{w} \sum_{q=-\Lambda/b}^{\Lambda/b}q^2
V(q)|\phi(q,w)|^2, \nonumber \\
\end{eqnarray}
where I dare to leave the velocity in the Gaussian part. Note that
I need not the renormalization of the velocity in the case $\beta >1$.
After the scale transformation, I obtain the eqs.
\begin{eqnarray}
 \frac{dg}{dl}&=&0 \nonumber \\
 \frac{d}{dl}(\frac{v}{K})&=&\frac{gA}{2\pi} \nonumber \\
 \frac{d}{dl}(\frac{1}{vK})&=&0,
\end{eqnarray}
where $A$ is the constant appearing in (\ref{longwave}).
I see that $K$ and the velocity $v$
is renormalized instead of the no renormalization of $g$.
The forward scattering become relevant through $K,v$ and drive the 
system away from the TL fixed point. Note that this result 
holds irrespective of any filling $k_{F}$.
From these eqs., the size dependences of $v$ and $K$ are given by
\begin{eqnarray}
v(b) 
%&=& \sqrt{ {\rm const.} + g\; {\rm const.} \ln b} 
&\sim & \sqrt{\ln L} 
\nonumber \\
K(b) &\sim & 1/\sqrt{\ln L}. \label{vKtendency}
\end{eqnarray}
The velocity diverges weakly for long distances, 
which is consistent with the estimations of $v=\frac{dw}{dq}$ from
the behaviours (\ref{longwave}). 
\subsection*{$\beta = 3$}
I use 
$V(q)= A+B q^{2} \ln q + C q^{2}+\cdots $ in the behaviours (\ref{longwave}).
The g term of (\ref{wavespaction}) is
\begin{eqnarray}
\sum_{w}\sum^{\lambda /b}_{q=-\lambda /b}
q^{2}(g_{1}(0) q^{2} \ln q+g_{2}(0) q^{2}), 
\end{eqnarray}
where the couplings $g_{1}(0)$ and $g_{2}(0)$ 
are defined by $g B$ and $g C$ respectively. 
%The couplings $g_{1}$ and $g_{2}$ are renormalized 
For the scale transformation (\ref{sctrans}), the g term is changed 
to 
\begin{eqnarray}
\sum_{w}\sum^{\lambda}_{q=-\lambda} q^{2}
[g_{1}(0)q^{2} \ln q /b^{2}+(g_{2}(0)/b^{2}+g_{1}(0)q^{2} \ln q /b^{2}))].
\end{eqnarray}
Thus I obtain 
\begin{eqnarray}
g_{1}(b) &=& \frac{1}{b^{2}}g_{1}(0) \nonumber \\
g_{2}(b) &=& g_{1}(0) \frac{1}{b^{2}} \ln \frac{1}{b}+
g_{2}(0)\frac{1}{b^{2}}. 
\end{eqnarray}
By $l=ln b$, 
I write this as
\begin{eqnarray}
\frac{d g_{1}(l)}{d l} &=& -2 g_{1}(l) \nonumber \\
\frac{d g_{2}(l)}{d l} &=& -2 g_{2}(l)-g_{1}(l). 
\end{eqnarray}
%I write the most relevant contribution in the main part of this chapter. 
\subsection*{$\beta > 3$}
This case is same as eq. (\ref{RGlngrng}) putting $\beta = 3$.
\section*{2. CFT in the TL liquid with LRI}
%\subsection*{$\beta >1$ irrelevant case}
The Hamiltonian in the finite strip from the action (\ref{action}) is 
\begin{eqnarray}
H &=& H_{\rm TL}+
%+g\int_{-L/2}^{L/2}
g\int_{D}
d\sigma_{1}d\sigma_{2}\partial_{\sigma_{1}}\phi(\sigma_{1})
\partial_{\sigma_{2}}\phi(\sigma_{2})
V(|\sigma_{1}-\sigma_{2}|)\theta(|\sigma_{1}-\sigma_{2}|-\alpha_{0}),
\label{fundhamil}
\end{eqnarray}
where $H_{\rm TL}$ is TL liquid and 
$D$ means the region 
$D=\{|\sigma_{1}-\sigma_{2}| \leq L, 
-L/2 \leq \sigma_{1},\sigma_{2}\leq L/2
\}$. I introduce
the step function $\theta(x)$ to avoid the ultra violet divergences
which come from $V(x)$ and the operator 
product expansion of $\partial_{\sigma}\phi(\sigma)$.
For the small perturbation $g$ the ground state energy
$E_{g}$ varies as
\begin{eqnarray}
  E^{'}_{g}- E_{g}&=& 
g\int_{D} d\sigma_{1}d\sigma_{2}V(|\sigma_{1}-\sigma_{2}|)
<0|\partial_{\sigma_{1}}\phi(\sigma_{1})\partial_{\sigma_{2}}\phi(\sigma_{2})|0> \theta(|\sigma_{1}-\sigma_{2}|-\alpha_{0})
\nonumber \\
  &=&-\frac{g}{4}\int_{D} d\sigma_{1}d\sigma_{2}V(|\sigma_{1}-\sigma_{2}|)
[ <0|\partial_{w_{1}}\varphi(w_{1})\partial_{w_{2}}\varphi(w_{2})|0> 
\nonumber \\
  & & + <0|\partial_{\bar{w}_{1}}\bar{\varphi}(\bar{w}_{1})
\partial_{\bar{w}_{2}} \bar{\varphi}(\bar{w}_{2})|0> ]_{\tau_{1}=\tau_{2}=0}
\theta(|\sigma_{1}-\sigma_{2}|-\alpha_{0}),
\end{eqnarray}
where I introduce the coordinates $w=\tau+i\sigma$ (
$-L/2 \leq \sigma \leq L/2$, $-\infty < \tau < \infty$)
and $|0>$ is the ground state of $H_{\rm TL}$. 
From the characters of the Gaussian part (TL liquid part) I can 
separate as $\phi(\sigma,\tau)=
\varphi(w)+\bar{\varphi}(\bar{w})$ and derive
$<0|\partial_{\bar{w}_{1}}\bar{\varphi}(\bar{w}_{1})
\partial_{w_{2}} \varphi(w_{2})|0>=0$.
The content of the brackets is modified as follows:
%\widetext
\begin{eqnarray}
&&[ <0|\partial_{w_{1}}\varphi(w_{1})\partial_{w_{2}}\varphi(w_{2})|0>+
<0|\partial_{\bar{w}_{1}}\bar{\varphi}(\bar{w}_{1})
\partial_{\bar{w}_{2}} \bar{\varphi}(\bar{w}_{2})|0> 
]_{\tau_{1}=\tau_{2}=0} \nonumber \\
&=& \frac{K}{4}[  (\frac{2\pi}{L})^{2\Delta} \frac{z_{2}}{z_{1}}
\frac{1}{(1-\frac{z_{2}}{z_{1}})^2} 
+ (\frac{2\pi}{L})^{2\bar{\Delta}} \frac{\bar{z}_{2}}{\bar{z}_{1}}
\frac{1}{(1-\frac{\bar{z}_{2}}{\bar{z}_{1}})^2}  
]_{\tau_{1}=\tau_{2}=0} \nonumber \\
&=& -\frac{K}{4}(\frac{2\pi}{L})^{2} \frac{1}{2\sin^2 \frac{\pi 
(\sigma_{1}-\sigma_{2})}{L} },
\end{eqnarray}
where I transform the correlation function
$<\partial_{z_{1}} \tilde{\varphi}(z_{1})\partial_{z_{2}}
\tilde{\varphi}(z_{2})>
=\frac{K}{4(z_{1}-z_{2})^2}$
in $\infty \times \infty$ $z$ plane
to that in the strip $w$ thorough $z=\exp{\frac{2\pi w}{L}}$. 
At present case $\partial_{w}\tilde{\varphi}(w)$
($\partial_{\bar{w}}\bar{\tilde{\varphi}}(\bar{w})$)
have the spin $s=1(-1)$ and conformal dimension $\Delta=1$($\bar{\Delta}=1$). 
Hence I obtain
\begin{eqnarray}
  E^{'}_{g}-E_{g}&=& 
%\frac{g}{4} (\frac{2\pi}{L})^{2} L 
%\int_{-L/2}^{L/2}d x
%V(|x|)\frac{1}{\sin^2 \frac{\pi x}{L}}
%\theta(|x|-\alpha_{0}) \nonumber \\
%&=& \frac{g(2\pi)^{2}}{4} \int_{-1/2}^{1/2}dx^{'}V(|Lx^{'}|)
%\frac{1}{\sin^2 \pi x^{'}}\theta(L|x^{'}|-\alpha_{0}) \nonumber \\
%&=&
\frac{g K \pi^{2} }{4} (\frac{\pi}{L})^\beta  \int_{-1/2}^{1/2}dx^{'}
\frac{1}{(\sin \pi |x^{'}|)^{\beta} \sin^2 \pi x^{'}  }
\theta(|x^{'}|-\frac{\alpha_{0}}{L}),
\end{eqnarray}
where I impose the periodic boundary 
condition and use the interaction potential $V(x)=
1/(\frac{L}{\pi}\sin(\frac{x \pi}{L}))^\beta$. 
Putting $\epsilon=\alpha_{0}/L$ for convenience, I give 
the differential of the integral part: 
\begin{eqnarray}
&& \frac{\partial}{\partial \epsilon}
\int_{-1/2}^{1/2}dx^{'}
\frac{1}{(\sin \pi |x^{'}|)^{\beta} \sin^2 \pi x^{'} }
\theta(|x^{'}|-\epsilon)
=-\frac{2}{(\sin \pi |\epsilon|)^{\beta} \sin^2 \pi \epsilon}.
\end{eqnarray}
After integrating the Taylor expansion about $\epsilon$ of this
quantity, I obtain
\begin{eqnarray}
\int_{-1/2}^{1/2}dx^{'}
\frac{1}{(\sin \pi |x^{'}|)^{\beta}}
\frac{1}{\sin^2 \pi x^{'}}\theta(|x^{'}|-\epsilon)
&=&{\rm const.}+\frac{2}{\pi}[\frac{(\pi \epsilon)^{-\beta-1}}{\beta+1}
+\frac{\beta+2}{6(\beta-1)}  (\pi \epsilon)^{-\beta+1} \nonumber \\
&&
+\frac{1}{\beta-3} \{ \frac{1}{120}(\beta+2)-\frac{1}{72}(\beta+1)(\beta+2)\}
(\pi \epsilon)^{-\beta+3} \nonumber \\
&&
+O((\pi \epsilon)^{-\beta+5})], 
\label{betadepend}
\end{eqnarray}
where $\beta \neq$ odd.
Therefore I can write the corrections in the form:
\begin{eqnarray}
E^{'}_{g} - E_{g}&=& \frac{gK}{2}[ \frac{{\rm A(\beta)}}{L^\beta}
+{\rm B(\beta)}L
+\frac{{\rm C(\beta)}}{L}+
\frac{{\rm D(\beta)}}{L^3}+O(\frac{1}{L^5})].\label{appscale}
\end{eqnarray}
%where $A(\beta)$, $B(\beta)$, C and D are the functions of $\beta$.
%
% Though 
%it is not easy to determine $A$, 
%Comparing with (\ref{betadepend}), 
Here $B(\beta)$, $C(\beta)$ and $D(\beta)$ are given by
\begin{eqnarray}
 B(\beta) &=& \frac{\alpha_{0}^{-\beta-1}}{1+\beta} \nonumber \\
 C(\beta) &=& \frac{\pi^{2}(2+\beta)}{6(\beta-1)} \alpha_{0}^{-\beta+1} \\
 D(\beta) &=& \frac{\pi^{4}}{\beta-3}
\{ \frac{1}{120}(\beta+2)-\frac{1}{72}(\beta+1)(\beta+2) \} 
\alpha_{0}^{3-\beta}. \nonumber
\label{Db}
\end{eqnarray}
I can obtain $A(\beta)$ by evaluating the above integral numerically. 
The result is shown in Fig. \ref{integral2}(a).
%Hence the C term contributes to the deviations of central charge.
%Note that these terms
%penetrate into the present corrections even if $\beta=0$. 
%Thus I think that the intrinsic contributions of the long-range interactions are 
%\begin{eqnarray}
%E^{'}_{g} - E_{g}&=& 
%g[ \frac{{\rm A}}{L^\beta}+B\frac{1}{L^3}+O(\frac{1}{L^5})].
%\end{eqnarray}
%This result is consistent with the spectrum analysis
%and the RG results for $\beta >1$ \footnote{If the B and C term 
%in eq. (C. 7) 
%give the important contributions to
%the excitations, 
%the TLL is broken by such the contributions. This
%is discrepant with the spectrum results and the RG arguments.
%Hence I assume that the B and C contributions
%in eq. (C. 7) are not intrinsic. Especially I should 
%regard the B term nonuniversal bulk constants}.

For $\beta$= odd, there exists the logarithmic
correction instead of the eq. (\ref{appscale}). 
The results for respective $\beta$ are
\begin{equation}
E^{'}_{g} - E_{g}=
\begin{cases}
g[ \frac{{\rm A}}{L}+{\rm B}L
+\frac{{\rm C}}{L}\ln \frac{1}{L}+D\frac{1}{L^3}+O(\frac{1}{L^5})]
&\text{\rm for $\beta=1$} \\
g[ \frac{{\rm A}}{L^3}+{\rm B}L
+\frac{{\rm C}}{L}
+D\frac{1}{L^3}\ln\frac{1}{L}
+O(\frac{1}{L^5})]
&\text{\rm for $\beta=3$} \\
g[ \frac{{\rm A}}{L^5}+{\rm B}L
+\frac{{\rm C}}{L}
+D\frac{1}{L^3}+E\frac{1}{L^5}\ln\frac{1}{L}+O(\frac{1}{L^7})]
&\text{\rm for $\beta=5$} \\
\cdots .
\end{cases}
\label{appscale2}
\end{equation}   
The C terms in eqs. (\ref{appscale}) and (\ref{appscale2})
contribute to deviation of the central charge.
%should be renormalized to
%TLL, because
The present LRI
inevitably contains the contribution
from the short range interaction: $\delta (x)$.
The C term does not come from the short range types of interactions
because the vacuum expected value 
$\langle (\partial_{x}\phi)^2 \rangle $ vanishes.
Thus C term is intrinsic in the present system with
the LRI under periodic boundary condition.
Because the velocity is not
renormalized as I have seen from the renormalization group eqs., the C term contributes
to the deviations of the
effective central charge from the ground state energy.

Next I derive the corrections for the energy of the excited state:
\begin{eqnarray}
E_{n}^{'}-E_{n}&=&
g
\int_{D} d\sigma_{1}d\sigma_{2}V(|\sigma_{1}-\sigma_{2}|)
<n|\partial_{\sigma_{1}}\phi(\sigma_{1})\partial_{\sigma_{2}}\phi(\sigma_{2})|n>\theta(|\sigma_{1}-\sigma_{2}|-\alpha_{0})
\nonumber \\
&=& 
g
\int_{D} d\sigma_{1}d\sigma_{2}V(|\sigma_{1}-\sigma_{2}|)
\sum_{\alpha}
<n|\partial_{\sigma_{1}}\phi(\sigma_{1})|\alpha><\alpha|
\partial_{\sigma_{2}}\phi(\sigma_{2})|n>
\theta(|\sigma_{1}-\sigma_{2}|-\alpha_{0})
\nonumber \\
%&=& 
%g\sum_{\alpha}C_{n j \alpha} C_{\alpha j n} (\frac{2\pi}{L})^{2}
%\int_{D} d\sigma_{1}d\sigma_{2}
%V(|\sigma_{1}-\sigma_{2}|)
%e^{2\pi i(s_{n}-s_{\alpha})(\sigma_{1}-\sigma_{2})/L}
%\theta(|\sigma_{1}-\sigma_{2}|-\alpha_{0}) \nonumber \\
%\nonumber \\
%&=& 4 g \sum_{\alpha}C_{n j \alpha} C_{\alpha j n}
%(\frac{2\pi}{L})^{2} L\int_{0}^{L/2}dx V(|x|)
%\cos \frac{2\pi}{L}(s_{n}-s_{\alpha})x
%\;\;\theta(|x|-\alpha_{0}) \\
&=& 4 g \sum_{\alpha}C_{n j \alpha} C_{\alpha j n} \frac{(2\pi)^2}{L^{\beta}}
\int_{0}^{1/2}dy \frac{1}{(\sin \pi |y|)^\beta}
\cos 2\pi(s_{n}-s_{\alpha})y\;\;\theta(|y|
-\frac{\alpha_{0}}{L}),\label{excLdep}
\end{eqnarray}
where I use the results by Cardy \cite{Cardy4}:
\begin{eqnarray}
<n|\phi(\sigma)|\alpha>&=&C_{nj\alpha}(\frac{2\pi}{L})^{x_{j}}e^
{\frac{2\pi i(s_{n}-s_{\alpha})\sigma}{L}}.
\end{eqnarray}
Here j means $\partial \phi$ and $|n>$ is the excited state of $H_{\rm TL}$.
I can derive the size dependence of eq. (\ref{excLdep})
from likewise treatments as the ground state.
After taking the derivative about $1/L$, I expand about $1/L$. 
Integrating them, I obtain 
\begin{eqnarray}
&& E_{n}^{'}-E_{n} \nonumber \\
&=&
\begin{cases}
16 \pi^{2} g \sum_{\alpha}C_{n j \alpha} C_{\alpha j n}
\{ \frac{A(s_{n}-s_{\alpha},\beta)}{L^\beta}+
\frac{B(\beta )}{L} 
+\frac{C(s_{n}-s_{\alpha},\beta)}{L^3}
+\frac{D(s_{n}-s_{\alpha},\beta)}{L^5}+O(\frac{1}{L^7})\}
\;\; 
& \text{$\beta \neq {\rm odd }$} \\
16 \pi^{2} g \sum_{\alpha}C_{n j \alpha} C_{\alpha j n}
\{ \frac{A(s_{n}-s_{\alpha})}{L}+B\frac{1}{L}\ln \frac{1}{L}
+C(s_{n}-s_{\alpha})\frac{1}{L^3}
+D(s_{n}-s_{\alpha})\frac{1}{L^5}+O(\frac{1}{L^7})\}
& \text{$\beta=1$} \\
16 \pi^{2} g \sum_{\alpha}C_{n j \alpha} C_{\alpha j n}
\{ \frac{A(s_{n}-s_{\alpha})}{L^3}+B\frac{1}{L}
+C(s_{n}-s_{\alpha})\frac{1}{L^3} \ln \frac{1}{L} 
+D(s_{n}-s_{\alpha})\frac{1}{L^5}+O(\frac{1}{L^7})\}
&\text{$\beta=3$} \\
16 \pi^{2} g \sum_{\alpha}C_{n j \alpha} C_{\alpha j n}
\{ \frac{A(s_{n}-s_{\alpha})}{L^5}+B\frac{1}{L}
+C(s_{n}-s_{\alpha})\frac{1}{L^3} 
+D(s_{n}-s_{\alpha})\frac{1}{L^5}\ln \frac{1}{L} 
+O(\frac{1}{L^7})\}
&\text{$\beta=5$}  \\
\cdots ,
\end{cases}
\nonumber \\
\label{fscaleapp}
\end{eqnarray}
where $B$ are the constant independent of $s_{n},s_{\alpha}$.
Here for $\beta \neq$ odd, $B(\beta)$, $C(s_{n}-s_{\alpha},\beta)$ and
$D(s_{n}-s_{\alpha},\beta)$ are given by
\begin{eqnarray}
B(\beta) &=& \frac{1}{(\alpha_{0} \pi)^{\beta-1}\pi (\beta-1)} \nonumber \\
C(s_{n}-s_{\alpha},\beta) &=& \frac{1}{(\alpha_{0} \pi)^{\beta -3}}
[\frac{\beta}{6}-2(s_{n}-s_{\alpha})^{2}]
\frac{1}{\pi( \beta -3)} \\
D(s_{n}-s_{\alpha},\beta) &=& 
\frac{1}{(\alpha_{0} \pi)^{\beta -5}}
[(-\frac{(s_{n}-s_{\alpha})^{2}}{3}+\frac{1}{180})\beta
+\frac{\beta^{2}}{72}+\frac{2(s_{n}-s_{\alpha})^{4}}{3}]
\frac{1}{\pi( \beta -5)} \nonumber
.
\end{eqnarray}
It is not straightforward to determine $A(s_{n}-s_{\alpha},\beta)$ generally. 
However about one particle excitation($s_n=0$), I can obtain $A(\beta,s_{\alpha})$, 
which is shown for some $s_{\alpha}$ in Fig. \ref{integral2}(b). 
Actually further consideration
about the operator product expansion leads $s_n=s_\alpha=0$ (see section 4.).

I refer to the $O(1/L)$ dependences. 
These are due to the fact that 
the LRI includes the short range type interaction. 
Actually I can derive the same form 
\begin{eqnarray}
\frac{g}{L}\sum_{\alpha}C_{n j \alpha} C_{\alpha j n}
\end{eqnarray}
as an ordinary finite scaling by replacing as $V(|x|)\theta(|x|-\alpha_{0}) \rightarrow \delta(x)$.
As $(\partial \phi)^2$ is a part of the TL liquid, the $O(1/L)$ 
term can be erased under subtracting such the contributions first.
Thus the $O(1/L)$ term is not intrinsic.

Summarizing the discussions in this appendix, 
I can prove that the Hamiltonian (\ref{fundhamil}) is described by 
$c=1$ CFT for $\beta > 1$ 
in the excitation energy. However 
the effective central charge from the ground state 
depends on the interaction
and deviates from 1.
I find the nontrivial behaviors when $\beta = \rm odd$, which 
corresponds to the integer points of the modified Bessel function
as appearing in the behaviours (\ref{longwave}).
\end{document}